\documentclass[prl, twocolumn, a4paper, showpacs]{revtex4}

\usepackage{amsfonts}
\usepackage{amsmath,amssymb}
\usepackage{graphicx}
\usepackage[small,bf]{caption}

\newcommand{\lb}[0] { \left( }
\newcommand{\rb}[0] { \right) }
\newcommand{\beqs} { \begin{eqnarray} }
\newcommand{\eeqs} { \end{eqnarray} }
\newcommand{\bsub} { \begin{subequations} }
\newcommand{\esub} { \end{subequations} }
\newcommand{\nn} {\nonumber}
\newcommand{\ep}[0] { \epsilon }

\newcommand{\de}[0] { \delta }

\newcommand{\EE}[2] {#1 \times 10^{#2}}

\newcommand{\rmax}{\, {\rm max}}

\newcommand{\sync}[0] { \rm sync}
\newcommand{\dyn}[0] { \rm dyn}
\newcommand{\dec}[0] { \rm dec}
\newcommand{\acc}[0] { \rm acc}

\newcommand{\gyr}[0] { \rm gyr}
\newcommand{\intsh}{ \rm int}
\newcommand{\iso}{ \rm iso}
\newcommand{\eff}{ \rm eff}
\newcommand{\br}{ \rm br}
\newcommand{\thr}{ \rm th}
\newcommand{\dif}{ \rm diff}
\newcommand{\SN}{ \rm SN}

\begin{document}

\title{Enhanced high-energy neutrino emission
from choked gamma-ray bursts due to meson and muon acceleration }
\author{Hylke B. J. Koers}
\affiliation{Service de Physique Th\'eorique, Universit\'e Libre de Bruxelles (U.L.B.),
CP225, Bld. du Triomphe, B-1050 Bruxelles, Belgium}
\author{Ralph A. M. J. Wijers}
\affiliation{Anton Pannekoek Instituut, Kruislaan 403, 1098 SJ Amsterdam, The Netherlands}

\begin{abstract}
It has been suggested that a potentially large fraction of
supernovae could be accompanied by relativistic outflows
that stall below the stellar surface. In this letter we
point out that internal shocks that are believed to accelerate
protons to very high energies in these flows will also accelerate
secondary mesons and muons. As a result the neutrino spectrum from meson
and muon decay is expected to be much harder compared to previous
estimates, extending as a single
power law up to $\sim$$10^3$ TeV.
This greatly improves the detection
prospects.

\end{abstract}


%95.85.Ry Neutrino, muon, pion, and other elementary particles; cosmic rays 
%96.50.Pw Particle acceleration 
%98.70.Rz gamma-ray sources; gamma-ray bursts 
%97.60.Bw Supernovae

\maketitle

Based on the observational connection (see Ref. \cite{Woosley:2006fn} for a
review)
between gamma-ray bursts (GRBs) and supernovae (SNe), it has been hypothesized that a
sizable fraction of all SNe is accompanied by a relativistic outflow
similar to those that are believed to lie at the base of observed GRBs
\cite{Meszaros:2001ms}.
Initially
the flow accelerates by radiation pressure and ploughs through the pre-burst
stellar material. The flow may however be `choked'
below the stellar surface
when the central engine is not active for a sufficiently long time.
As electromagnetic radiation that is dissipated by these flows will be absorbed
by the stellar material, neutrinos are likely the only particles
that could indicate the existence of this phenomenon.

High-energy ($\gtrsim 1$ TeV) neutrinos arise predominantly in the decay of charged mesons and muons that are created in the interactions of shock-accelerated protons with
target protons or photons. In particular, neutrino emission due to
proton acceleration in internal shocks  in the relativistic flow 
has been studied in detail
\cite{Meszaros:2001ms, Razzaque:2003uv,
Razzaque:2004yv, Razzaque:2005bh,  Ando:2005xi}.
It has been argued that
the fluence of high-energy neutrinos in this scenario is suppressed 
because the mesons and muons lose a
substantial amount of energy before decay due to synchrotron emission and hadronic interactions 
\cite{Razzaque:2004yv, Ando:2005xi}. This energy loss 
strongly limits the neutrino detection prospects.
However, as we point out here, the mesons and muons
are also deflected and scattered  by the strong magnetic field 
on very short timescales. These particles
have ample time to cross the shock several times before decay
and hence they are 
subject to shock acceleration.
In this work we
investigate the observational consequences
of the acceleration of neutrino parent particles.
For concreteness we restrict ourselves to internal shocks
in choked GRB outflows. We expect however that the acceleration mechanism is quite general in GRBs, so that our results may affect  
estimates of neutrino fluxes in other scenarios as well.

\emph{The model --- } 
We consider a relativistic outflow with total energy
$E = 10^{52} E_{52}$, Lorentz factor $\Gamma_j = 10 \Gamma_{j,1}$ and
opening angle $\theta = 0.1 \theta_{-1}$
(we use the notation $Q = 10^x Q_x$ throughout this letter). 
Here $\theta$
parameterizes the combined effect of collimation
and relativistic beaming.
The  isotropic-equivalent burst energy is 
$E_{\iso} = 2 E / \theta^2  = \EE{2}{54} \textrm{ erg} \times
\theta_{-1}^{-2}  \, E_{52}$.
Following Refs. \cite{Razzaque:2004yv, Razzaque:2005bh,Ando:2005xi},
we assume that internal shocks occur in the flow
at a radius
$r_{\intsh} = 2 c \Gamma_j^2 \delta t = \EE{6}{11} \textrm{ cm}
\times \, \Gamma_{j,1}^2 \,  {\delta t}_{-1}$,
where $\delta t = 0.1 \de_{j,-1}  \, {\rm s}$  is the variability timescale of the central engine. 
The comoving proton density  at the internal shock radius is 
$ n'_{p} = E_{\iso} / 4 \pi r_{\intsh}^2 \Gamma_j^2 m_p c^3 t
= 10^{19} \textrm{ cm}^{-3} \times 
\Gamma_{j,1}^{-6} \, \theta_{-1}^{-2} \, E_{52} \, t_{1}^{-1} \,  {\delta t}_{-1}^{-2} $,  where $t = 10 t_1 \, {\rm s}$ is the burst duration
(quantities in the comoving frame are denoted with a prime)
and $m_p$ is the proton mass.
The large proton density gives rise to a very high Thomson optical depth
$\tau \gtrsim 10^6$. This implies that
synchrotron photons that are emitted by shock-accelerated electrons will thermalize. 
The number density of photons at
the internal shock radius is
$n'_{\gamma}= 0.33 (\ep_e U'/ \hbar c)^{3/4}  
= \EE{2}{23} \textrm{ cm}^{-3} \times
\Gamma_{j,1}^{-9/2} \, \theta_{-1}^{-3/2} \,  (\ep_e E)_{51}^{3/4} \,  t_1^{-3/4} \, {\delta t}_{-1}^{-3/2} $, where
 $U' =  E_{\iso} / 4 \pi r_{\intsh}^2 \Gamma_j^2 c t$
is the comoving energy density and $\ep_e
= 0.1 \ep_{e,-1}$ denotes the
fraction of the total energy in the flow that is transfered to the
thermal photon distribution.
The magnetic field strength at the internal shock radius is 
$ B' = \EE{2}{8} \, {\rm G} \times  \Gamma_{j,1}^{-3} \theta_{-1}^{-1} (\ep_B E)_{52}^{1/2}
t_1^{-1/2} {\de t}_{-1}^{-1} $,
where $\ep_B = 0.1 \ep_{B,-1}$  denotes the
ratio of electromagnetic energy to the total energy in the flow.

We adopt the SN rate parameterization presented in Ref.
\cite{Porciani:2000ag}. The SN rate
within proper distance $d_p$ can be approximated with
$ \dot{N}_{\rm SN} = \EE{4}{2}$ year$^{-1} \times d_{p,2}^3$, where
$d_p = 100 \, d_{p,2}$ Mpc. This 
estimate is within $\sim$$30$\% for $d_p \lesssim 500$ Mpc.
Due to collimation of the relativistic outflow,
the rate of observable choked GRBs associated with these SNe
is $\dot{N}_{\rm CGRB} = \theta^2 \dot{N}_{\rm SN} /2 = 2 \textrm{ yr}^{-1} \times \xi_{\SN}  \theta_{-1}^2 d_{p,2}^3$, where $\xi_{\SN} \leq 1$ is the fraction of SNe
that is  endowed with the type of outflows considered in this work.

\emph{Proton acceleration --- }
We assume that internal shocks  accelerate a fraction of the protons in
the flow to high energies (see also below). We express the energy 
spectrum of accelerated protons as
$d N_p / d E'_p = \xi_p  E_{\iso} {(p-1)} \Gamma_j^{-1} {E'_p}^{-p} (m_p c^2)^{p-2}$,
where  $\xi_p = 0.01 \xi_{p,-2}$ denotes the fraction of 
shock-accelerated protons to all nucleons in the flow, and
$p$ is the shock-acceleration power-law index.
Theoretical studies \cite{Achterberg:2001rx} indicate that 
$p \simeq 2.3$, but recent observations 
suggest that $p$ could vary \cite{Starling:2007kk}. To keep the discussion general we consider 
here the range $p = 2.0 \ldots 2.6$.
The proton acceleration timescale
is equal to 
$t'_{p, \, \acc}  = E'_p / q c B' \Gamma'_s =
\EE{6}{-11} \, {\rm s} \times E'_{p,0} {\Gamma'}_{s,1}^2 \theta_{-1}
(\ep_B E)_{51}^{-1/2} \delta t_{-1} t_1^{1/2}$,
where $E'_{p} = 1 E'_{p,0} \, {\rm TeV}$ denotes the proton energy
and $\Gamma'_s = 10 \Gamma'_{s,1}$ is the Lorentz factor of the shock
\cite{Achterberg:2001rx}.
We take $\Gamma'_s = \Gamma_j$ since the variation in Lorentz factors between two
subsequent shells of material $\Delta \Gamma_j \sim \Gamma_j$.
The maximum proton energy may be limited both by
energy losses and by
the finite shock lifetime. The lifetime is roughly equal 
to the dynamical timescale
$t'_{\dyn}  = r_{\intsh} / c \Gamma_j =
2 \, {\rm s} \times \Gamma_{j,1} \delta t_{-1}$.
The dominant proton energy-loss mechanisms are
synchrotron radiation, photopion production and proton-proton ($pp$) collisions. The synchrotron energy-loss timescale
is  $t'_{p, \, \sync}  =  (6 \pi m_p^4 c^3) / (\sigma_{\rm T} m_e^2 {B'}^2 E'_p) = 0.1 \, {\rm s} \times {E'}^{-1}_{p,0} \Gamma_{j,1}^6 \theta_{-1}^2
(\ep_B E)_{51}^{-1} \delta t_{-1}^2 t_1$,
where we assume that the protons are relativistic.
The energy-loss timescale due to $pp$ collisions is
$t'_{p, \, pp}  = 1 / (c K_{pp} \sigma_{pp} (1-\xi_p) n'_p )  = 
10^{-4} \, {\rm s} \times  \Gamma_{j,1}^6 \theta_{-1}^2
E_{52}^{-1} \delta t_{-1}^2 t_1 $, where we assume that $\xi_p \ll 1$,
and we approximate the cross section for $pp$ collisions with
$\sigma_{pp} \simeq \EE{5}{-26}$ cm$^2$ and the fractional energy loss with $K_{pp} \simeq 0.5$. At center-of-mass energies well above the pion creation threshold we
approximate the proton-photon ($p \gamma$) cross section with
$\sigma_{p \gamma} \simeq 10^{-28}$ cm$^2$, and the fractional energy loss
to pion production with $K_{p \gamma \pi} \simeq 0.2$. In this regime 
the bulk of the photons
participates in photopion production so that we estimate
the photopion energy-loss timescale as
$ t'_{p,p \gamma \pi} \simeq 1/ (c \sigma_{p \gamma} K_{p \gamma \pi}
n'_{\gamma}) =\EE{9}{-6} \, {\rm s} \times \Gamma_{j,1}^{9/2} \theta_{-1}^{3/2} (\ep_e E)_{51}^{-3/4}   \delta t_{-1}^{3/2} t_1^{3/4} $.
Equating the acceleration timescale to the dynamical timescale and
the energy-loss timescales we find that, unless extreme values of the parameters
are invoked, the maximum proton energy $E'_{p,\rmax}$
is determined by synchrotron energy loss.
In this case
\beqs
E'_{p,\rmax}
= \EE{5}{4} \, {\rm TeV} \times \Gamma_{j,1}^{2} \theta_{-1}^{1/2}
(\ep_{B}  E)_{51}^{-1/4} {\de t}_{-1}^{1/2} t_1^{1/4} \, .
\eeqs
For protons with sufficient energy the optical depth for photopion
production $\tau_{p \gamma}  \simeq 2 \sigma_{p \gamma} n'_\gamma  r_{\intsh}/\Gamma_j
= \EE{2}{6} \times \Gamma_{j,1}^{-7/2} \theta_{-1}^{-3/2}
( \ep_{e} E)_{51}^{3/4} {\de t}_{-1}^{-1/2} t_1^{-3/4} $
is always larger than
unity (unless $\Gamma_j$ is very large). On the other hand,
the optical depth for $pp$ interactions
$\tau_{p p} \simeq \sigma_{pp} n'_p r_{\intsh} / \Gamma_j
= \EE{3}{4} \times \Gamma_{j,1}^{-5} \theta_{-1}^{-2}
  E_{52} {\de t}_{-1}^{-1} t_1^{-1}$
is less than unity when $\Gamma_j \gtrsim 80$. 
In this case 
protons with energy below the photopion threshold may traverse the jet relatively
unhindered
and impact directly on the jet head or the stellar material, 
which could have
interesting observational consequences.
Here we assume that
protons lose all their energy in the outflow and that 
20\% of this energy is transfered to secondary mesons.

\emph{Meson and muon acceleration --- }
The secondary mesons and muons are created in a strongly magnetized environment. The magnetic field deflects and scatters the particles through electromagnetic interactions on timescales comparable to the
gyration timescale
$t'_{x, \, \gyr} = \ep'_x / q c B'
= \EE{6}{-10} \, {\rm s} \times \ep'_{x,0} \Gamma_{j,1}^3 \theta_{-1} (\ep_B E)_{51}^{-1/2} {\delta t}_{-1} t_1^{1/2} $,
where $x$ denotes either $\mu$ (muon), $\pi$ (pion) or $K$ (kaon), 
and $\ep'_x = 1 \, \ep'_{x,0} \, {\rm TeV}$ is the particle energy.
The comoving decay time is given by
$t'_{x, \, \dec} = \tau_x \ep'_x / m_x c^2 = 
\EE{2}{-2} \, {\rm s} \, (\EE{2}{-4}\, {\rm s}, \, \EE{2}{-5} \, {\rm s})  \times  \ep'_{x,0}$  for
muons (pions, kaons), where
$\tau_x$ denotes the proper decay time
and $m_x$ denotes the mass. Since 
the decay time is much longer than the gyration time,
the particles have ample time to
be deflected and scattered by the magnetic field.
This allows them to cross the shock repeatedly, thereby
gaining energy in a stochastic way.
This is essentially the same mechanism of
shock acceleration that applies to protons. A necessary condition for this mechanism
to work is that the acceleration timescale
$ t'_{x, \, \acc} =  t'_{x, \, \gyr} / \Gamma'_s \ll t'_{x, \, \dec}$.
Notice that the ratio $ t'_{x, \, \acc} / t'_{x, \, \dec}$ is independent of 
the particle energy, so that
there is no intrinsic maximum energy to the acceleration process.

For stable particles
the energy spectrum due to shock acceleration can be approximated with a power law with index
$p = 1 - {\rm ln} ( \mathcal{P}_{\rm ret}) / {\rm ln} \chi$, where
$\mathcal{P}_{\rm ret} \simeq 0.5$ is the return probability (i.e. the probability that a particle completes a full cycle of two shock crossings), and $\chi \simeq 1.6-2.0$ is the average 
fractional energy gain per cycle \cite{Achterberg:2001rx}. 
For unstable particles the energy spectrum at decay (which
determines the energy spectrum of the daughter particles)
is, in principle, expected to be softer than the energy spectrum of 
accelerated stable particles because fewer particles complete many cycles.
The effect of particle decay can be accounted for
by adopting the return probability
$\tilde{\mathcal{P}}_{\rm ret} =  \mathcal{P}_{\rm ret}  - 
\mathcal{P}_{\rm dec}$, where $ \mathcal{P}_{\rm ret} $ is
the return probability for stable particles and
$\mathcal{P}_{\rm dec} \simeq t'_{x, \acc} / t'_{x, \dec}$ is the probability
that a particle decays during one cycle. 
The acceleration timescale of charged particles in the internal shock
environment is
$t'_{x, \, \acc} = \ep'_x / q c B' \Gamma'_s
= \EE{6}{-11} \, {\rm s} \times \ep'_{x,0} \Gamma_{j,1}^2 \theta_{-1} (\ep_B E)_{51}^{-1/2} {\delta t}_{-1} t_1^{1/2} $.
Since $t'_{\dec} \gg t'_{\acc}$ over a wide range of parameters,
$\mathcal{P}_{\rm dec} \ll  \mathcal{P}_{\rm ret}$ and we
expect that the mesons and muons are accelerated
to a power law with index $p \simeq 2.3$ that extends to
the maximum energy determined by the shock lifetime or by energy losses.
We find that, both for mesons and muons, the maximum energy ${\ep'_x}^{\hspace{-0.07cm} \max}$
is  determined by synchrotron losses.
Equating the acceleration timescale 
to the  synchrotron energy-loss timescale
$t'_{x, \, \sync}  = t'_{p, \, \sync}  (m_x / m_p)^4 $, we estimate:
\bsub
\label{eq:eppiKmumax}
\beqs
{\ep'_{\mu}}^{\hspace{-0.07cm} \max} =
\EE{6}{2} \, {\rm TeV} \times \Gamma_{j,1}^{2} \theta_{-1}^{1/2} (\ep_{B} E)_{51}^{-1/4} {\de t}_{-1}^{1/2} t_1^{1/4} \, ; \\ 
{\ep'_{\pi}}^{\hspace{-0.07cm} \max}=  10^3 \, {\rm TeV}
 \times \Gamma_{j,1}^{2} \theta_{-1}^{1/2} (\ep_{B} E)_{51}^{-1/4} {\de t}_{-1}^{1/2} t_1^{1/4}\, ; \\
{\ep'_{K}}^{\hspace{-0.07cm} \max} = 10^4 \, {\rm TeV}
 \times \Gamma_{j,1}^{2} \theta_{-1}^{1/2} (\ep_{B} E)_{51}^{-1/4} {\de t}_{-1}^{1/2} t_1^{1/4}\, .
\eeqs
\esub

\emph{Neutrino fluence --- }
We denote the average neutrino
multiplicity per proton
with $\mathcal{M}_{p \nu (x)}$ 
and the average fraction of the proton energy
that is transferred to the neutrino
with $\xi_{p \nu (x) }$.
Here $x$ labels the intermediate
particle (muon, pion, kaon).
Following Refs.
\cite{Razzaque:2005bh,Ando:2005xi},
we assume that a proton transfers 20\% of its energy 
per collision to the secondary mesons.
Furthermore we take the average pion (kaon) multiplicity per proton interaction
equal to 1 (0.10). The branching ratio of pion (kaon) decay to a 
muon and a neutrino is virtually unity (0.63) and
the neutrino receives $\sim$$0.25$ (0.50) of the meson energy.
A muon transfers $\sim$0.33 of its energy to each of
two daughter neutrinos.
Hence  $\mathcal{M}_{p \nu (\mu)} = 2, \,  \mathcal{M}_{p \nu (\pi)} = 1, \, \mathcal{M}_{p \nu (K)} = 0.06$;
 $\xi_{p \nu (\mu)} = 0.05, \, \xi_{p \nu (\pi)} = 0.05, \, \xi_{p \nu (K)} = 0.1$.

The differential neutrino fluence in the observer frame
(all flavours combined; neutrinos and antineutrinos combined)  can be expressed as follows:
\beqs
\label{eq:int:nufluence1}
\Phi_{\nu(x)} (\ep_\nu) & = & \frac{1}{4 \pi d_p^2}
\frac{1}{\Gamma_j} \frac{\mathcal{M}_{p \nu (x)} }{\mathcal{\xi}_{p \nu (x)} }
\frac{d N_p}{d E_p} \\
\nn & =&  \tilde{\Phi}_{\nu(x)}  \Gamma_{j,1}^{p-2}
\theta_{-1}^{-2} E_{52}  \xi_{p,2}  d_{p,2}^{-2}
 \lb \frac{\ep_\nu}{1 \, {\rm TeV}} \rb^{-p} \, ,
\eeqs
where we assume that the redshift $z \ll 1$, and
\beqs
\label{eq:int:nutilde}
\tilde{\Phi}_{\nu(x)}& =& \EE{5.2}{-4} \, {\rm TeV}^{-1} {\rm cm}^{-2} \\
\nn & & \times \, \mathcal{M}_{p \nu (x)}
\lb \frac{\xi_{p \nu (x)}}{0.05} \rb^{p-1}
(p-1) (\EE{4.7}{-4})^{p-2} \, .
\eeqs
From eqs. \eqref{eq:eppiKmumax}, we find that 
the maximum neutrino energy  in the observer frame
is:
\beqs
\nn \ep^{\max}_{\nu (\mu)} & = &
\EE{2}{3} \, {\rm TeV} \times \Gamma_{j,1}^{3} \theta_{-1}^{1/2}
(\ep_{B} E_{51})^{-1/4}  {\de t}_{-1}^{1/2} t_1^{1/4} \, ; \\
\nn \ep^{\max}_{\nu (\pi)} & = &
\EE{3}{3} \, {\rm TeV} \times \Gamma_{j,1}^{3} \theta_{-1}^{1/2}(\ep_{B} E_{51})^{-1/4}  {\de t}_{-1}^{1/2} t_1^{1/4}  \, ; \\
\nn \ep^{\max}_{\nu (K)} & = &
\EE{6}{4} \, {\rm TeV} \times \Gamma_{j,1}^{3} \theta_{-1}^{1/2} (\ep_{B} E_{51})^{-1/4}  {\de t}_{-1}^{1/2} t_1^{1/4} \, ,
\eeqs
where we
take the neutrinos to be isotropic in the 
comoving frame.

\emph{Detection prospects --- }
Based on preliminary results presented in Ref. \cite{Desiati:2006qc} we 
conservatively approximate
the effective area of IceCube for
muon neutrinos with:
\beqs
\nn 
A_{\eff} (\ep_{\nu}) = \left\{ 
\begin{array}{c l}
\EE{5.0}{2} \, {\rm cm}^2 \,  \times \lb \ep_{\nu,0} \rb^{1.7} &  (\ep^0_{\thr} < \ep_\nu < \ep_{\br} ) \\
\EE{1.5}{5} \, {\rm cm}^2 \, \times \lb \ep_{\nu,0} \rb^{0.3} & (\ep_\nu > \ep_{\br} )
\end{array} 
\right. \,,
\eeqs
where $\ep_{\thr} = 0.1$ TeV denotes the detector threshold energy,  $\ep_{\br} = 60$ TeV 
is a break energy, and $ \ep_\nu = 1 \, \ep_{\nu,0} \, {\rm TeV}$. 
As the neutrino flavour ratio at the source is roughly
${\nu_e : \nu_\mu : \nu_\tau} \simeq
{1:2:0}$,
the expected neutrino flavour ratio 
at the detector is $\simeq {1: 1:1}$
due to neutrino oscillations over very large distances
\cite{Learned:1994wg}.
Hence we approximate the fluence of muon neutrinos at the detector
with
\beqs
\label{eq:int:nufluence2}
\Phi_{\nu_{\mu}} (\ep_\nu) \simeq \frac{1}{3} \lb \Phi_{\nu(\mu)} (\ep_\nu)  + \Phi_{\nu(\pi)} (\ep_\nu) 
+ \Phi_{\nu(K)} (\ep_\nu) 
\rb \, .
\eeqs
We estimate the number
of muon-neutrino interactions $N_{\nu_\mu}$ in IceCube by multiplying
the muon-neutrino fluence 
with the effective area and integrating over $\ep_\nu$ to find:
\beqs
\label{eq:det:Nnu}
N_{\nu_\mu} \simeq \tilde{\Phi}_{\nu_\mu}  \tilde{N}_{\nu_\mu} 
 \Gamma_{j,1}^{p-2} \theta_{-1}^{-2} E_{52}   \xi_{p,2} d_{p,2}^{-2}
 \, ,
\eeqs
where $\tilde{\Phi}_{\nu_\mu} = (\tilde{\Phi}_{\nu (\mu)} + \tilde{\Phi}_{\nu (\pi)}
+ \tilde{\Phi}_{\nu (K)})/3 $, and
\beqs
\label{eq:det:Psi}
\nn \tilde{N}_{\nu_\mu} =
 \frac{\EE{9}{3}\, {\rm TeV \, cm}^{2}}{60^{p-2}} \lb \frac{1- x_{\thr}^{2.7-p}}{2.7-p}
- \frac{1- x_{\max}^{1.3-p}}{1.3-p} \rb \, .
\eeqs
Here we use the shorthand notation
$x_{\thr} \equiv \ep_{\thr} / \ep_{\br}$ and
$x_{\max} \equiv \ep_{\nu}^{\max} / \ep_{\br}$, and
assume that $ \ep_{\nu}^{\max}> \ep_{\br}$
and that $1.3<p<2.7$. In the limit 
that $\ep_{\nu}^{\max} \gg \ep_{\br}$,
we find that
$\tilde{N}_{\nu_\mu} = \EE{2}{4}$ ($\EE{8}{3}$, $\EE{4}{3}$) TeV cm$^2$ for
$p=2.0$  ($2.3$, $2.6$). 
Combining this with eqs. \eqref{eq:int:nutilde} and \eqref{eq:det:Nnu},
we predict $N_{\nu_\mu} = 
14 $ ($0.63$, $0.038$) $\Gamma_{j,1}^{p-2}  \theta_{-1}^{-2} E_{52}  \xi_{p,-2} d_p^{-2}
$ muon-neutrino interactions in
IceCube for $p=2.0$ ($2.3$, $2.6$).
Hence, for model parameters similar to those
adopted in this study,
a choked GRB at 100 Mpc
could be observed by IceCube provided that the shock-acceleration
index $p$ is not too large. For reference
values of the other parameters the detection of one neutrino requires
$p \lesssim 2.3$.

\begin{figure}
\includegraphics[width=5.5cm, angle=270]{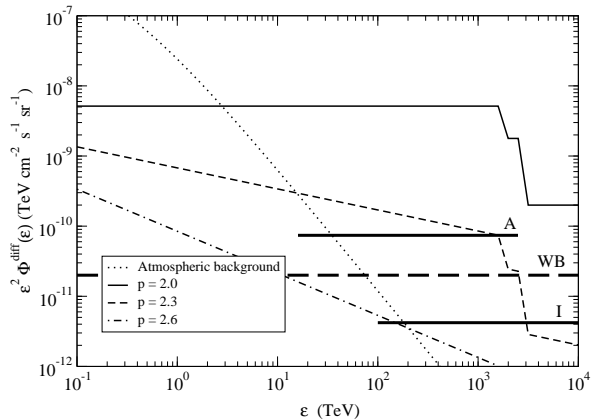} 
\caption{\label{figure:diff}
Diffuse muon-neutrino flux at Earth for  three values of the power-law index $p$ and
reference values of the other parameters.  Also shown are the atmospheric neutrino
background, the AMANDA-II (A) and IceCube (I)
upper limits, and the Waxman-Bahcall bound (WB).}
\end{figure}

The diffuse flux (per sterad) due to unresolved choked GRBs can be estimated with
\beqs
\Phi^{\dif}_{\nu_\mu} (\ep_{\nu})
 = \frac{\xi_{\SN}   \theta^2}{8 \pi} \int_0^{\infty} dz  \lb \frac{dV}{dz} \rb \dot{n}_{\SN}
\Phi_{\nu_\mu} (\ep_{\nu}) \, ,
\eeqs
where 
$\xi_{\SN}$ denotes the fraction of SNe that is accompanied by a choked GRB,
$V$ is the comoving volume, $z$ is the redshift, and
$\dot{n}_{\SN}$ in the SN rate per unit volume for which
we adopt the parameterization
given in Ref. \cite{Porciani:2000ag} (see also Ref.
\cite{Razzaque:2005bh}).
We find that:
\beqs
\label{eq:Phidiff}
\Phi^{\dif}_{\nu_\mu} (\ep_{\nu})
 = \tilde{\Phi}^{\dif}_{\nu_\mu} \Gamma_{j,1}^{p-2} E_{52} \xi_{p,2} 
\xi_{\SN} \lb \frac{\ep_{\nu}}{1 {\rm TeV}}\rb^{-p} \, ,
\eeqs
where
$\tilde{\Phi}^{\dif}_{\nu_\mu} =
\EE{9}{-6} \, \tilde{\Phi}_{\nu_\mu} \, {\rm sr}^{-1} $.
In figure \ref{figure:diff} we have plotted our estimate \eqref{eq:Phidiff} of the diffuse muon-neutrino flux for 
three values of $p$
and reference values of the other parameters.
Also shown are the 90\%  confidence
level upper limits of the AMANDA-II
\citep{Achterberg:2007qp}
and IceCube (3 year) \citep{Ahrens:2003ix}
experiments, the atmospheric neutrino background
with the parameterization used in 
Ref. \cite{Razzaque:2005bh}, and the Waxman-Bahcall bound
\cite{Waxman:1998yy}.
As can be seen in the figure, the existing limit from the AMANDA-II experiment
is already constraining the parameter space of choked GRBs.
For $p=2.0$ (2.3), we find that $E_{52} \xi_{p,2} \xi_{\SN} \lesssim 10^{-2}$ ($1$). 
In this regime the predicted diffuse flux is above the Waxman-Bahcall bound, which applies to optically thin sources.
When $p=2.6$ the expected diffuse
flux is below the detector sensitivity for reference values of the other parameters.
IceCube will be able to put more stringent constraints on
the model parameters. A more detailed analysis of the detection prospects
is beyond the scope of this work. We note however
that a visible SN counterpart, which would provide evidence in favour of
our model, would strongly reduce the neutrino 
background. It may also be feasible to use neutrino detectors to
initiate a SN search.

\emph{Conclusions --- }
\label{sec:concanddisc}
In this letter we have found that muons, pions, and kaons created by
$pp$ and $p \gamma$ interactions in the internal shocks 
of choked GRB outflows are also accelerated by these shocks, essentially
in the same way as protons and electrons are. Although
this acceleration mechanism
appears to be unavoidable under the circumstances believed to be
present in these flows, it was (to the best of our knowledge)
not considered before.
Due to meson and muon acceleration the resulting neutrino spectrum
is expected to be dominated by neutrinos from muon decay and to
follow a single power law with index
$p \simeq 2.3$ up to a maximum energy $\sim$$10^3$ TeV
(and larger than $10^4$ TeV for the subdominant contribution of
neutrinos from kaon decay).
This is in contrast to
previous estimates that predict  spectral breaks at  energies
$\lesssim 1$ TeV  for neutrinos
from meson decay and virtually no high-energy neutrinos
from muon decay \cite{Razzaque:2004yv, Razzaque:2005bh,Ando:2005xi}.
The relatively hard neutrino spectrum
strongly increases the detection prospects. In fact, the current AMANDA-II limit
on the diffuse neutrino background is already mildly constraining the model
parameters. The upcoming IceCube neutrino detector will be in a good position to test the
model and, possibly, to observe neutrino emission from single choked GRBs.
We have estimated that a single
choked GRB at 100 Mpc aimed toward Earth 
will result in $14 $ ($0.63$, $0.038$) muon-neutrino interactions in IceCube for
$p = 2.0$ ($2.3$, $2.6$) and reference values of the other parameters.
The rate of choked GRBs emitting neutrinos toward Earth within 100 Mpc
may be as large as a few per year.

An important caveat in our results is that the model relies on the
existence of internal shocks that occur
at a radius $r_{\intsh}$ due to variability in the flow.
It is however
not clear whether these shocks can indeed develop while the jet is traversing
the star and has not yet created a low-density funnel (Thomas Janka,
private communication). In contrast to this,
a forward shock and a reverse shock seem unavoidable in the interaction
of the relativistic outflow with the stellar environment. Also
internal shocks are expected to occur
behind the forward shock as it propagates through the star.
Neutrino production
in these shocks will be studied in a forthcoming
publication.
We have also assumed that a fair fraction of the secondary
mesons makes its way to the shock after being produced. This has to be verified in a more detailed study. We
expect that this assumption is best justified
at high proton energies, where the proton mean free path is 
not much larger than the meson gyroradius.

Shock acceleration of mesons and muons
may be a fairly general phenomenon in GRBs.
The necessary condition that the acceleration timescale
is smaller than the decay timescale is fulfilled
when $ B \Gamma_j > 5$ ($\EE{6}{2}$, $\EE{5}{3}$) G for muons 
(pions, kaons), which is easily achieved in GRBs.
Hence our results may also affect the estimates for neutrino emission
from successful GRBs.

\emph{Acknowledgments --- }
H.K. would like to thank Dimitrios Giannios,
Thomas Janka and Peter Tinyakov for useful discussions.
H.K. is supported by Belgian Science Policy under IUAP VI/11
and by IISN.
H.K. acknowledges financial support by FOM during the development
of this work. 
R.W. acknowledges support from NWO through vici grant
639.043.302.

\end{document}